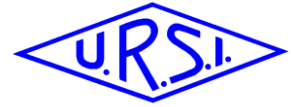

# VIPER: A Plasma Wave Detection Instrument onboard Indian Venus Orbiter Spacecraft

Vipin K. Yadav

Space Physics Laboratory (SPL), Vikram Sarabhai Space Centre (VSSC), Indian Space Research Organization (ISRO), Thiruvananthapuram 695022, Kerala, India

## Abstract

Plasma waves are observed in almost all the solar system objects - planets, their satellites, comets, Sun, interplanetary medium, etc. The planetary ionospheres are capable of sustaining plasma waves which are observed there and play an important role in the ionospheric dynamics by propagating energy across different space regions and provide acceleration to particles to attain high energies for transportation. The study of planetary plasma waves also provides information on the solar wind – planet interaction, the energy distribution in the ionospheric plasma of that planet, etc.

Venus does not possess a global magnetic field unlike Earth. Thesolar EUV radiation ionizes the neutrals and generates a plasma environment around Venus which can sustain plasma waves. Very few attempts are made to observe all plasma waves that can exist around Venus and that too with instruments having a limited dynamic range such as with PVO (Pioneer Venus Orbiter) and Venus Express. However, there are some other plasma waves which can exist around Venus but are yet to be observed.

ISRO is planning to send an orbiter mission to Venus in near future where a suit of instruments named VIPER (Venus Ionospheric Plasma wavE detectoR) is onboard to observe Venusian plasma waves. The plasma wave observations around Venus and VIPER onboard ISRO's Venus Orbiter Spacecraft are discussed in this paper.

## 1. Introduction

Venus is a terrestrial planet which does not possess a global magnetic field. Thesolar radiation (dominantly EUV) interacts with the dense Venus atmosphere and ionizes large number of neutral atoms and molecules and the ionosphere gets generated. The magnetic field observed around Venus is a result from the solar wind interaction with the upper atmosphere of the Venus. The induced magnetic field of Venus varies ~ 50 to 150 nT.

The first plasma wave detection around Venus was carried out by PVO in 1978 with a suit of instruments such as magnetometer [1, 2], electric dipole [3] and electron temperature probe, [4].

The electric dipole had an effective length of 0.75 m formed by a pair of electric field sensors having wire circles of diameter 10.5 cm. The electric field sensor had four frequency channels: 100 Hz, 730 Hz, 5.4 kHz and 30 kHz. The automatic gain control amplifiers had a rise time of 50 ms and a decay time of about 500 ms. The Orbiter Electron Temperature Probe (OETP) was designed to measure the plasma parameters in Venus ionosphere - electron ($n_e$) and ion ($n_i$) plasma density, electron plasma temperature ($T_e$) and the spacecraft potential ($V_{sc}$). It consisted of two cylindrical sensors – one placed radially at the end of a 1 m long boom and the other placed axially at a distance of 0.4 m away from the spacecraft with a common electronics unit [4]. The Venus Express in 2005 from ESA also carried two triaxial magnetometer sensors to measure the magnetic field in the Venus ionosphere with one sensor mounted directly on the spacecraft and the other at the end of a 1 m boom. The magnetometer sensors have a dynamic range from ± 32.8 to ± 8388.6 nT with 128 Hz cadence [5]. Table 1 lists different plasma wave detection instruments which have flown onboard various missions to Venus for measurements.

**Table 1**

| Mission | Year | Agency/ Country | Plasma wave Instrument | Other Plasma Instruments |
|---|---|---|---|---|
| Mariner-10 | 1974 | NASA | Magnetometer | Plasma Ion Probe |
| Pioneer Venus Orbiter (PVO) | 1978 | NASA | Orbiter Electric Field Detector (OEFD), Magnetometer (OMAG), Orbiter Electron Temperature Probe (OETP) | Orbiter Ion Mass Spectrometer (OIMS), Orbiter Plasma Analyzer (OPA), Orbiter Retarding Potential Analyzer (ORPA), Orbiter Radio Occultation |
| Venera 11/12 | 1978 | USSR | Magnetometer | ---- |
| Venus Express | 2005 | ESA | Magnetometer | ASPERA-4 |

PVO measured the electron plasma density varies from a maximum of $5 \times 10^5$ cm$^{-3}$ at 150 km to about $10^4$ cm$^{-3}$ at 900 km whereas the electron temperature has a variation between $1100^o$ K at 150 km to $8000^o$ K at about 900 km. The ion temperature is less than electron temperature at the same altitude at $500^o$ K at 150 km to $1300^o$ K at 900 km. The typical plasma and magnetic field parameters around Venus [6, 7] are summarized in Table 2.

**Table 2**

| Plasma parameter | Range |
|---|---|
| Electron number density, $n_e$ (cm$^{-3}$) | $3 \times 10^2 - 5 \times 10^5$ |
| Electron temperature, $T_e$ (eV) | 0.1 – 1.7 |
| Ion number density, $n_i$ (cm$^{-3}$) | $10 - 2 \times 10^5$ |
| Ion temperature, $T_i$ (eV) | 0.02 – 0.3 |
| Background magnetic field, B (nT) | 50 – 165 |

## 2. Venus Plasma Wave Observations

PVO was the first best equipped spacecraft to observe plasma waves near the Venus. Its electric field sensors observed a strong and highly variable solar wind interaction with the Venusian ionosphere. PVO observed electromagnetic whistler waves in 100 Hz, lower hybrid waves in 30 kHz range [8] and electron plasma oscillations in 20-54 kHz frequency range [9]. Galileo during its flyby of Venus observed electrostatic Langmuir waves [10]. Venus Express detected proton cyclotron waves in Venus ionosphere with a set of fluxgate magnetometers onboard [11, 12]. Mirror mode waves were also observed in the induced magnetosphere of Venus by Venus Express [13, 14]. Apart from the above mentioned waves, ion acoustic waves (IAW) with frequency $f_{IAW} \approx 100$ Hz are proposed to be present in the Venus ionosphere as a consequence of the ion acoustic beam instability due to $O^+$ ions [15]. Magnetosonic waves having frequencies between 40-50 mHz (in the spacecraft frame) are observed by Venus Express [16]. A review on plasma waves around Venus is given elsewhere [17]. A summary of various plasma wave observed in the ionosphere of Venus is given in Table 3.

**Table 3**

| Observed Plasma Waves | Spacecraft | Instruments | Frequency |
|---|---|---|---|
| Electron plasma oscillations | PVO, Galileo | OEFD, MAG | ≈ 50 kHz |
| IAW | PVO | OEFD, OMAG, OETP | 730 Hz, 5.4 kHz |
| Whistler waves, | | | = < 100 Hz |
| Lower hybrid waves | | | 30 kHz |
| Langmuir waves | Galileo | PWS, MAG | 200 Hz - 7 kHz |
| Proton cyclotron waves | Venus Express | MAG | = < 0.5 Hz |
| Mirror mode waves | | | 30 – 300 mHz |
| Magnetosonic waves | | | 40 – 50 mHz, < 100 mHz |

Table 4 lists all the plasma waves that are expected to be present around Venus and are to be observed by VIPER.

**Table 4**

| Plasma Waves/oscillations | Expected frequency |
|---|---|
| Hydromagnetic waves | 40 – 50 mHz, < 100 mHz |
| Mirror mode waves | 30 – 300 mHz |
| Proton cyclotron waves | = < 0.5 Hz |
| Whistler waves | = < 100 Hz |
| Ion cyclotron waves | 19 – 625 Hz |
| Mix-mode waves | 50 – 1000 Hz |
| Electron cyclotron waves | 1.4 – 4.5 kHz |
| Ion acoustic waves | 730 Hz, 5.4 kHz |
| Langmuir waves | 200 Hz - 7 kHz |
| Lower hybrid waves | 30 kHz |
| Electron plasma oscillations | ≈ 50 kHz |

Table 5 list the plasma waves that can exist around Venus but are yet to be observed.

**Table 5**

| Un-observed Plasma Waves | Expected frequency | Importance |
|---|---|---|
| Electron cyclotron waves | 1.4 – 4.5 kHz | Electron transport from Venus |
| Ion cyclotron waves | 19 – 625 Hz | Particle loss from the Venusian ionosphere |
| Mix-mode waves | 50 – 1000 Hz | Venus ionospheric heating |

A hand-made schematic distribution of plasma waves around Venus is shown in Figure 1.

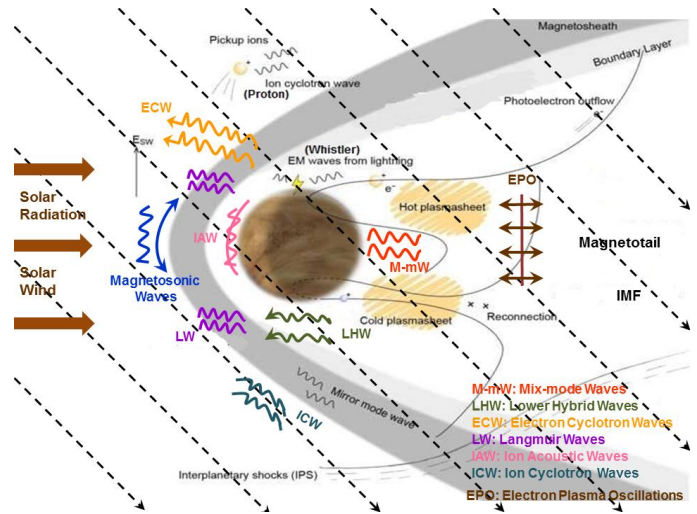

**Figure 1.** The distribution of plasma waves around Venus (Not to scale in altitude).

It is to be noted that there are only couple of attempts made to observe all plasma waves that can exist in the Venus ionosphere and that too with instruments having a limited dynamic range. However, the limited instrumental capability onboard PVO and Venus Express shows that a number of plasma waves exist in the ionosphere of Venus.

There are some plasma waves which may exist in the ionosphere of Venus but are not detected such electron and ion cyclotron waves, mix-mode waves, etc. Although Langmuir, ion acoustic, whistler and proton cyclotron waves are detected by earlier Venus missions, it shall be prudent to make an attempt again to observe these plasma waves if the dynamic range of the measuring instrument makes it possible so as to verify the earlier measurements and to get some additional scientific information if possible.Electron plasma frequency ($f_{pe}$) & subsequently electron plasma (Langmuir) wave frequency ($f_{EPW}$) can be computed as follows: ($n_e \approx 10^3$ - $5 \times 10^5$ cm$^{-3}$, $f_{pe}$ (min.) = 0.284 MHz; $f_{pe}$ (max.) = 6.35 MHz.

## 3. VIPER: Science Objectives

The proposed scientific instruments, in the form of a suit (VIPER) are – a customized Langmuir probe (LP) to measure the electron & ion number density and electron plasma temperature, a triaxial electric field sensor (EFS) to measure the oscillating plasma wave electric field, a triaxial search-coil magnetometer (SCM) to measure the oscillating plasma wave magnetic field around Venus and a triaxial fluxgate magnetometer (FGM) to measure the background magneticfield around Venus. All these sensors shall be mounted on two separate booms thereby keeping them away from the spacecraft to avoid the sheath and shielding effects.

The scientific objectives of VIPER are:

(i) To study the plasma phenomena around Venus by exploring it in the prevailing localized regions. For this purpose, the electron and ion plasma parameters are going to be measured *in-situ* with LP onboard the orbiter spacecraft around Venus.

(ii) To study of the interplanetary and induced magnetic field structure around Venus. For this purpose, the background magnetic field is going to be continuously measured *in-situ* with FGM onboard the orbiter spacecraft around Venus.

(iii) To detect and observe the plasma waves around Venus and to explore their role in modulating the plasma dynamics around Venus. For this purpose, the plasma wave frequency is going to be measured with EFS and SCM.

Table 6 lists the scientific objectives pertaining to the measurement parameters and sampling requirements.

**Table 6**

| Science Objectives | Measurement parameters |
|---|---|
| To explore the plasma environment around Venus and to study the plasma phenomena prevailing in localized regions. | Continuous measurement of plasma parameters such as electron and ion number density, electron and ion plasma temperature around the Venus. |
| Study of the interplanetary and induced magnetic field structure prevailing around Venus. | Continuous measurement of background magnetic field around the Venus. |
| Characterization of Venusian plasma waves and their role in modulating the plasma dynamics around Venus. | Continuous detection and measurement of the plasma wave electric and magnetic fields along with the localized plasma parameters around the Venus. |
| Study of various physical phenomena taking place around Venus such as the plasma energy distribution, plasma heating, particle acceleration and their escape. | Continuous measurements of plasma wave parameters and charge particle distribution measurements along with their number densities and energies around the Venus. Data from VIPER and other onboard instruments to be used. |

## 4. VIPER Plasma Wave Detection

The identification of plasma waves by VIPER is done after detecting the time varying (oscillating) electric field and magnetic field signals. Initially the plasma parameters such as electron and ion number density and electron and ion plasma temperature shall be estimated using the LP along with the background steady state which is going to be measured by the FGM as given in Table 7.

**Table 7**

| Instrument / Scientific Aim | Plasma/Field Parameter & Range | Measurements Required |
|---|---|---|
| Langmuir Probe (LP) Plasma parameter measurements | $n_e$; [$10^2$ - $10^6$ cm$^{-3}$] $T_e$; [0.1 - 1 eV] $n_i$ (bulk); [$10^2$ - $10^5$ cm$^{-3}$] $T_i$ (bulk); [0.06 – 0.4 eV] | Electron & Ion saturation currents, Complete I-V characteristics |
| Electric Field Sensor (EFS) Oscillating plasma wave electric field | $\omega$ (ES & EM) [1 Hz - 55 kHz]; $E_1$; [mV/m] | Wave electric field at different frequencies. |
| Fluxgate Magnetometer (FGM) Background magnetic field measurements | $B_0$; [1 – 300 nT] | $B_{x1}$, $B_{y1}$, $B_{z1}$ and $B_{x2}$, $B_{y2}$, $B_{z2}$ |
| Search-coil Magnetometer (SCM) Oscillating plasma wave magnetic field | $\omega$ (EM only); [1 Hz – 55 kHz] $B_1$; [1 – 300 nT] | Wave magnetic field at different frequencies. |

Here, $\omega$ is the angular frequency (= $2\pi f$) of the wave having frequency $f$, $E_1$ and $B_1$ are the time-varying wave electric and magnetic fields respectively. These plasma and field parameters are to be used to estimate the plasma characteristic frequencies such as the electron/ion plasma frequency, electron/ion cyclotron frequency and secondary plasma parameters as given in Table 8.

**Table 8**

| Plasma Parameters | Measured quantities / constants | Value |
|---|---|---|

| Electron cyclotron frequency, $\omega_c = 2\pi (e B_0) / m_e$ | $B_0$, $e$, $m_e$ | $2.8 \times 10^6 B_0$ Hz |
|---|---|---|
| Ion cyclotron frequency, $\Omega_c = 2\pi (Z e B_0) / m_i$ | $B_0$, $m_i$, $m_p$, $Z$ $\mu = m_i/m_p$ | $1.52 \times 10^3 Z \mu^{-1} B_0$ Hz |
| Electron plasma frequency, $\omega_{pe} = 2\pi (e^2 n_e / m_e \varepsilon_0)^{1/2}$ | $n_e$, $e$, $m_e$, $\varepsilon_0$ | $8.98 \times 10^3 n_e^{1/2}$ Hz |
| Ion plasma frequency, $\omega_{pi} = 2\pi (Z^2 e^2 n_i / m_i \varepsilon_0)^{1/2}$ | $Z$, $e$, $n_i$, $m_i$, $\varepsilon_0$ | $2.1 \times 10^2 Z \mu^{-1/2} n_i^{1/2}$ Hz |
| Electron thermal velocity, $v_{the} = (K_B T_e / m_e)^{1/2}$ | $K_B$, $T_e$, $m_e$ | $4.19 \times 10^5 T_e^{1/2}$ m/sec |
| Ion sound velocity, $v_s = (\gamma Z K_B T_e / m_i)^{1/2}$ | $k$, $T_e$, $m_i$, $Z$, $\gamma$, $\mu$ $\mu = m_i/m_p$, $\gamma = c_p/c_v$ | $9.79 \times 10^3 (\gamma Z T_e /\mu)^{1/2}$ m/sec |
| Ion thermal velocity, $v_{thi} = (k T_i / m_i)^{1/2}$ | $K_B$, $T_i$, $m_e$ | $9.79 \times 10^3 T_i^{1/2} \mu^{-1/2}$ m/sec |
| Debye length, $\lambda_D = (K_B T_e / e^2 n_e)^{1/2}$ | $K_B$, $T_e$, $n_e$, $e$ | $7.43 T_e^{1/2} n_e^{-1/2}$ m |
| Alfven velocity $v_A = B_0 / (\mu_0 \rho)^{1/2}$ | $B_0$, $\mu_0$, $\rho$, $\mu$ $\rho = n_i m_i$, $\mu = m_i/m_p$ | $2.18 \times 10^9 \mu^{-1/2} n_i^{-1/2} B_0$ m/sec |

Here, $B_0$ is the background magnetic field, $e$ is electronic charge, $m_e$ is electron mass, $Z$ is the atomic number, $m_i$ is ion mass, $m_p$ is proton mass, $c$ is velocity of light in free space, $\varepsilon_0$ is the permittivity of the free space, $K_B$ is the Boltzmann constant, $T_e$ is the electron temperature, $\gamma$ is the ratio of specific heats, $\mu_0$ is the permeability of the free space, $\rho$ is the ion mass density, $n_i$ is ion number density.

The estimated plasma parameters and measured magnetic field along with the frequency of the oscillating electric and magnetic field shall be fitted in the dispersion relations of various plasma waves that can exist around the Venus and the outcome shall be analysed for proper identification. For example, the dispersion relation of an electron plasma (or Langmuir) wave is $\omega^2 = \omega_p^2 + (3/2) k^2 v_{th}^2$. Here, the variables are $\omega$ (measured from EFS), $\omega_{pe}$ (estimated from $n_e$ measured by LP), $v_{the}$ (estimated from $T_e$ measured by LP) and $k$. For a given small range of $k$, the dispersion relation shall be satisfied and the existence of the wave shall be established.

## 5. References

1. C.T. Russell, et al., "On the search for an intrinsic magnetic field at Venus", Proc. 11th Lunar Planetary Science Conference, 1980, pp. 1897-1906

2. C.T. Russell, et al., "Pioneer Venus orbiter fluxgate magnetometer", *IEEE Transactions on Geoscience and Remote Sensing*, **GE-18**, 1, 1980, pp. 32-35, doi:10.1109/TGRS.1980.350256

3. Scarf, et al., "The Pioneer Venus Orbiter Plasma Wave Investigation", *IEEE Transactions on Geoscience and Remote Sensing*, **GE-18**, 1, 1980, pp. 36-38, doi:10.1109/TGRS.1980.350257

4. J.P. Krehbiel, et al., "Pioneer Venus Orbiter Electron Temperature Probe", *IEEE Trans. on Geoscience and Remote Sensing*, **GE-18**, 1, 1980, pp. 49-54, doi:10.1109/TGRS.1980.350260

5. T.L. Zhang, et al., "Magnetic field investigation of the Venus plasma environment: Expected new results from Venus Express", *Planetary Space Science*, **54**, 2006, pp. 1336-1343, doi:10.1016/j.pss.2006.04.018

6. J.G. Luhmann, et al., "Characteristics of the Mars-like limit of the Venus-Solar wind Interaction", *Journal of Geophysical Research*, **92**, A8, 1987, pp. 8545-8557, doi: 10.1029/JA092iA08p08545

7. S.J. Bauer, et al., "The Venus Ionosphere", *Advances in Space Research*, **5**, 11, 1985, pp. 233-267, doi:10.1016/0273-1177(85)90203-0

8. F.L. Scarf, et al., "Plasma Waves Near Venus: Initial Observations", *Science*, **203**, 1979, pp. 748-750, doi:10.1126/science.203.4382.748

9. G.K. Crawford, R.J. Strangeway, and C.T. Russell, "Electron Plasma Oscillations in the Venus Foreshock", *Geophysical Research Letters*, **17**, 11, 1990, pp. 1805-1808, doi:10.1029/GL017i011p01805

10. G.B. Hospodarsky, et al., "Fine structure of Langmuir waves observed upstream of the bow shock at Venus", *Journal of Geophysical Research*, **99**, A7, 1994, pp. 13,363-13,371, doi: 10.1029/94JA00868

11. M. Delva, et al., "First upstream proton cyclotron wave observations at Venus", *Geophysical Research Letters*, **35**, 2008, L03105, doi:10.1029/2007GL032594

12. M. Delva, et al., "Proton cyclotron waves in the solar wind at Venus", *Journal of Geophysical Research*, **113**, 2008, E00B06, doi:10.1029/2008JE003148

13. M. Volwerk, et al., "First identification of mirror mode waves in Venus' magnetosheath?", *Geophysical Research Letters*, **35**, 2008, L12204, doi:10.1029/2008GL033621

14. M. Volwerk, et al., "Mirror-mode-like structures in Venus' induced magnetosphere", *Journal of Geophysical Research*, **113**, 2008, E00B16, doi:10.1029/2008JE003154

15. J.D. Huba, "Generation of waves in the Venus mantle by the Ion acoustic beam instability", *Geophysical Review Letters*, **20**, 17, 1993, pp. 1751-1754, doi:10.1029/93GL01984

16. L. Shan, et al., "Transmission of large-amplitude ULF waves through a quasi-parallel shock at Venus", *Journal of Geophysical Research - Space Physics*, **119**, 2014, pp. 237-245, doi:10.1002/2013JA019396

17. Vipin K. Yadav, Plasma Waves around Venus and Mars, *IETE - Technical Review*, **38**, 6, 2021, pp. 622-661, doi:10.1080/02564602.2020.1819889

18. F.F. Chen, "Introduction to Plasma Physics and Controlled Fusion", Third Edition, Chapter 4 "Waves in Plasma", Section 4.6 'Ion Waves', 1984, pp. 91